\documentclass[aps,prd,onecolumn,groupedaddress,showpacs,nofootinbib,amssymb]{revtex4}
\usepackage{graphicx,bm,color}
\usepackage{amsmath}
\usepackage{amssymb}
\usepackage{amsfonts}

\newcommand{\be}{\begin{equation}}
\newcommand{\ee}{\end{equation}}
\newcommand{\bea}{\begin{eqnarray}}
\newcommand{\eea}{\end{eqnarray}}
\newcommand{\beaa}{\begin{eqnarray*}}
\newcommand{\eeaa}{\end{eqnarray*}}

\newcommand{\nn}{\nonumber \\}
\newcommand{\e}{\mathrm{e}}

\allowdisplaybreaks[4]

\begin{document}

\tolerance=5000

\title{Gravitational Waves in the Presence of Viscosity}

\author{Iver Brevik$^1$\footnote{E-mail address: iver.h.brevik@ntnu.no}
and Shin'ichi Nojiri$^{2,3}$\footnote{E-mail address: nojiri@gravity.phys.nagoya-u.ac.jp}
}

\affiliation{
$^1$ Department of Energy and Process Engineering,
Norwegian University of Science and Technology, Trondheim, Norway \\
$^2$ Department of Physics, Nagoya University, Nagoya
464-8602, Japan \\
$^3$ Kobayashi-Maskawa Institute for the Origin of Particles and
the Universe, Nagoya University, Nagoya 464-8602, Japan}

\begin{abstract}

We analyze gravitational waves propagating in an isotropic cosmic fluid endowed with a bulk viscosity
$\zeta$ and a shear viscosity $\eta$, assuming these coefficients to vary with fluid density $\rho$
as $\rho^\lambda$, with $\lambda=1/2$ favored by experimental evidence.
We give the general governing equation for the gravitational waves, and focus thereafter on two
examples.
The first concerns waves in the very late universe, close to the Big Rip, where the fate of the comic
fluid is dependent highly on the values of the parameters.
Our second example considers the very early universe, the lepton era; the motivation for this choice
being that the microscopical bulk viscosity as calculated from statistical mechanics is then at
maximum.
We find that the gravitational waves on such an underlying medium are damped, having a decay
constant equal to the inverse of the conformal Hubble parameter. 
Our results turn out to be in good agreement 
with other viscosity-based approaches. 

\end{abstract}


\maketitle

\section{Introduction}

In hydromechanics in general the introduction of viscosity coefficients means
that one is working to the first order deviations from thermal equilibrium.
The usual model is that the viscous stress tensor is taken to be proportional to
the symmetric strain tensor $\frac{1}{2}(u_{i,k}+u_{k,i})$.
There should accordingly be three viscosity coefficients, but since the case
of uniform rotation does not imply any friction forces the number of coefficients is
reduced to two.
They are the shear viscosity $\eta$ proportional to the strain tensor, and
the bulk viscosity $\zeta$ proportional to $\nabla \cdot \bm{u}$.
The shear viscosity thus relates to fluid velocity gradients, while the bulk viscosity
relates only to compression or dilution of the fluid.
Usually, $\eta $ is much bigger than $\zeta $.

Also in cosmology the use of viscosity coefficients has been considered.
An early extensive treatment is that of Weinberg \cite{Weinberg:1971mx,Weinberg:1972kfs}.
A recent review, covering both the early- and the late-time universe, is given by
Brevik {\it et al.}~\cite{Brevik:2017msy} and \cite{Bamba:2012cp}.
There are several research papers in this area; some of them are
Refs.~\cite{Brevik:2005bj,Padmanabhan:1987dg,Gagnon:2011id,
Elizalde:2014ova,Brevik:2014eya,Brevik:2015cya,Brevik:2017juz,
Brevik:2016kwq,Normann:2016jns,Normann:2016zby,Nojiri:2005sr,Capozziello:2005pa}.

If we specialize to the case of gravitational waves, it turns out that also here
the viscosity concept has attracted attention
\cite{Hawking:1966qi,x71,madore73,Goswami:2016tsu}.
When applying viscosity to cosmology, the following point needs however some attention.
Viscosity is basically a macroscopic thermodynamical concept, applicable when
when the free paths are much shorter than the wavelengths.
The underlying mechanism is statistical mechanics and the Boltzmann equation,
and viscosity coefficients are usually calculated via
the so-called Chapman-Enskog expansion.
This expansion is based on the physical idea that for times much larger than
the mean free time, the statistical distribution function develops with time only via
the ``slowly'' varying quantities such as the particle density, the (mean) fluid velocity,
and eventually also the temperature \cite{chapman52,vincenti65}.

Now, the wavelengths observed in the present universe lie between 300 and 15,000 km.
There is apparently nothing in interstellar space that may provide such a small structure.
This point has recently been emphasized by Flauger and Weinberg \cite{Flauger:2017ged}.
In spite of such fundamental issues we will yet apply the viscosity idea in our
developments below.
The nature of the underlying medium, the dark matter, is after all not well known,
and not very much can be said about its local structure.
Moreover, the hydrodynamic model enables us to give a coherent formal view of
the evolution of the universe, both in its early and in its late epoch when even less
is known about the microstructure of the cosmic fluid.

In this paper we consider propagation of gravitational waves in a viscous isotropic fluid.
In the presence of bulk and/or shear viscosity there exist massless spin 2 waves,
which are influenced by the viscosity, and can increase or decay in amplitude dependent on
the parameters.
Dissipation, meaning transfer of energy into heat, is present under varying circumstances.
In the next section we review the general perturbed Einstein equations which describe
gravitational waves.
Then, we apply the formalism to the viscous fluid, in general both with a bulk viscosity
$\zeta$ and a shear viscosity $\eta$ present.
Both of them are modeled to be related to the energy density $\rho$ as $\rho^{1/2}$,
as this form appears to be most suitable when comparing with experiments.
We also present some numerical estimates.
We then go on to consider two examples: in Sec. VI we examine gravitational waves near
the future Big Rip singularity, distinguishing between cases when amplitudes are growing
and when they are decreasing.
In Sec. VII we consider a second example from the very early universe, from the lepton
epoch, especially from the instant of neutrino decoupling ($10^{10}~$K), when the bulk
viscosity as calculated from statistical mechanics was at maximum.
We give the governing equation for gravitational waves on a ``slowly varying'' geometric
background, based upon an experimentally favored bulk viscosity, and find an approximate
expression for the decay parameter for the amplitude.

\section{Brief Review of Gravitational Waves}

Before we consider the gravitational wave in the viscous fluid, we review
the propagation of gravitational waves in a general medium.
The equation of the gravitational wave is given by the perturbation of the Einstein
equation.
In the Einstein equation, not only the curvature but the energy-momentum tensor also
depends on the metric and therefore the variation of the energy-momentum tensor gives
a non-trivial contribution for the propagation of the gravitational wave.
We should note that the metric dependence of the energy-momentum tensor
highly depends on the model.
Even if different models give identical expansion history of the universe, the metric dependence
of the energy-momentum tensor and therefore the propagation of the gravitational wave is different
in the models (see \cite{Capozziello:2017vdi,Nojiri:2017hai,Bamba:2018cup} for examples).

In general, the perturbed Einstein equation is given by
\begin{align}
\label{GRGW1}
0 =& \frac{1}{2\kappa^2}\left(- \frac{1}{2}\left(\nabla^{(0)}_\mu
\nabla^{(0)\, \rho} \delta g_{\nu\rho}
+ \nabla^{(0)}_\nu \nabla^{(0)\, \rho} \delta g_{\mu\rho} - \Box^{(0)} \delta g_{\mu\nu}
 - \nabla^{(0)}_\mu \nabla^{(0)}_\nu \left(g^{(0)\, \rho\lambda}\delta g_{\rho\lambda}\right)
\right. \right. \nn
& \left. - 2R^{(0)\, \lambda\ \rho}_{\ \ \ \ \ \nu\ \mu}\delta g_{\lambda\rho}
+ R^{(0)\, \rho}_{\ \ \ \ \ \mu}\delta g_{\rho\nu}
+ R^{(0)\, \rho}_{\ \ \ \ \ \nu}\delta g_{\rho\mu} \right) \nn
& \left. + \frac{1}{2} R^{(0)} \delta g_{\mu\nu}
+ \frac{1}{2}g^{(0)}_{\mu\nu} \left( -\delta g_{\rho\sigma} R^{(0)\, \rho\sigma}
+ \nabla^{(0)\, \rho} \nabla^{(0)\, \sigma} \delta g_{\rho\sigma}
 - \Box^{(0)} \left(g^{(0)\, \rho\sigma}\delta g_{\rho\sigma}\right) \right) \right)
+ \frac{1}{2} \delta T_{\mathrm{matter}\, \mu\nu} \, .
\end{align}
By multiplying $g^{(0)\, \mu\nu}$ with (\ref{GRGW1}), we obtain
\begin{equation}
\label{GRGW1B}
0 = \frac{1}{2\kappa^2}\left( \nabla^{(0)\, \sigma}\nabla^{(0)\, \rho} \delta g_{\sigma\rho}
 - \Box^{(0)} \left(g^{(0)\, \rho\lambda}\delta g_{\rho\lambda}\right)
+ \frac{1}{2} R^{(0)} \left(g^{(0)\, \rho\lambda}\delta g_{\rho\lambda}\right)
 - 2 \delta g_{\rho\sigma} R^{(0)\, \rho\sigma} \right)
+ \frac{1}{2} \delta T_\mathrm{matter} \, .
\end{equation}
We choose the following gauge condition
\begin{equation}
\label{GRGW2}
0 = \nabla^{(0)\, \mu} \delta g_{\mu\nu} \, .
\end{equation}
Then Eq.~(\ref{GRGW1}) reduces to
\begin{align}
\label{GRGW3}
0 =& \frac{1}{2\kappa^2}\left(- \frac{1}{2}\left( - \Box^{(0)} \delta g_{\mu\nu}
 - \nabla^{(0)}_\mu \nabla^{(0)}_\nu \left(g^{(0)\, \rho\lambda}\delta g_{\rho\lambda}\right)
 - 2R^{(0)\, \lambda\ \rho}_{\ \ \ \ \ \nu\ \mu}\delta g_{\lambda\rho}
+ R^{(0)\, \rho}_{\ \ \ \ \ \mu}\delta g_{\rho\nu}
+ R^{(0)\, \rho}_{\ \ \ \ \ \nu}\delta g_{\rho\mu} \right) \right. \nn
& \left. + \frac{1}{2} R^{(0)} \delta g_{\mu\nu}
+ \frac{1}{2}g^{(0)}_{\mu\nu} \left( -\delta g_{\rho\sigma} R^{(0)\, \rho\sigma}
 - \Box^{(0)} \left(g^{(0)\, \rho\sigma}\delta g_{\rho\sigma}\right) \right) \right)
+ \frac{1}{2} \delta T_{\mathrm{matter}\, \mu\nu} \, ,
\end{align}
and Eq.~(\ref{GRGW1B}) to
\begin{equation}
\label{GRGW4}
0 = \frac{1}{2\kappa^2}\left(
 - \Box^{(0)} \left(g^{(0)\, \rho\lambda}\delta g_{\rho\lambda}\right)
+ \frac{1}{2} R^{(0)} \left(g^{(0)\, \rho\lambda}\delta g_{\rho\lambda}\right)
 - 2 \delta g_{\rho\sigma} R^{(0)\, \rho\sigma} \right)
+ \frac{1}{2} \delta T_\mathrm{matter} \, .
\end{equation}
By assuming the FRW space-time with flat spatial part (\ref{FRWmetric}),
we have
\begin{align}
\label{E2}
& \Gamma^t_{ij}=a^2 H \delta_{ij}\, ,\quad \Gamma^i_{jt}=\Gamma^i_{tj}=H\delta^i_{\ j}\, ,
\quad \Gamma^i_{jk}=\tilde \Gamma^i_{jk}\, ,\quad
R_{itjt}=-\left(\dot H + H^2\right)a^2\delta_{ij}\, ,\quad
R_{ijkl}= a^4 H^2 \left(\delta_{ik} \delta_{lj}
 - \delta_{il} \delta_{kj}\right)\, ,\nn
& R_{tt}=-3\left(\dot H + H^2\right)\, ,\quad
R_{ij}=a^2 \left(\dot H + 3H^2\right)\delta_{ij}\, ,\quad
R= 6\dot H + 12 H^2\, , \quad
\mbox{other components}=0\, .
\end{align}
Then $(t,t)$, $(i,j)$, $(t,i)$ components of (\ref{GRGW3}) have the following forms,
\begin{align}
\label{GRGW5}
0 =& \frac{1}{2\kappa^2}\left( \frac{1}{2} \Box^{(0)} \delta g_{tt}
+ \frac{1}{2} \partial_t^2 \left(g^{(0)\, \rho\lambda}\delta g_{\rho\lambda}\right)
+ \frac{1}{2} \Box^{(0)} \left(g^{(0)\, \rho\sigma}\delta g_{\rho\sigma}\right) \right. \nn
& \left. - \frac{1}{2} \left(\dot H - H^2\right) \left(g^{(0)\, ij}\delta g_{ij}\right)
 - \frac{3}{2} \left(\dot H - H^2\right) \delta g_{tt} \right)
+ \frac{1}{2} \delta T_{\mathrm{matter}\, tt}
\, ,\\
\label{GRGW6}
0 =& \frac{1}{2\kappa^2}\left( \frac{1}{2} \Box^{(0)} \delta g_{ij}
+ \frac{1}{2} \left( \partial_i \partial_j - H \delta_{ij} \partial_t \right)
\left(g^{(0)\, \rho\lambda}\delta g_{\rho\lambda}\right)
 - \frac{1}{2} g^{(0)}_{ij} \Box^{(0)} \left(g^{(0)\, \rho\sigma}\delta g_{\rho\sigma}\right)
+ \frac{1}{2} \left( \dot H + H^2 \right) g^{(0)}_{ij} \delta g_{tt} \right. \nn
& \left. + 2 \left( \dot H + H^2 \right) \delta g_{ij}
 - \frac{1}{2} g^{(0)}_{ij} \left( \dot H + H^2 \right) \left(g^{(0)\, kl}\delta g_{kl}\right)
\right) + \frac{1}{2} \delta T_{\mathrm{matter}\, ij} \, ,\\
\label{GRGW7}
0 =& \frac{1}{2\kappa^2}\left( \frac{1}{2} \Box^{(0)} \delta g_{ti}
+ \frac{1}{2} \nabla^{(0)}_t \nabla^{(0)}_i \left(g^{(0)\, \rho\lambda}\delta g_{\rho\lambda}\right)
+ \left(2 \dot H + 4 H^2\right) \delta g_{ti} \right)
+ \frac{1}{2} \delta T_{\mathrm{matter}\, ti} \, .
\end{align}

\section{Characteristics of a Viscous Fluid}

We now briefly review the basic properties of viscous fluid denoting, as already mentioned,
the shear viscosity by $\eta$ and the bulk viscosity by $\zeta$.
In the coordinates comoving with the viscous fluid the four velocity $\left( U^\mu \right)$
of the fluid is given by $U^0=1, U^i=0$.
The projection tensor, which corresponds to the spatial
directions perpendicular to $U^\mu$, is given by
\begin{equation}
\label{1}
h_{\mu\nu}=g_{\mu\nu}+U_\mu U_\nu\, .
\end{equation}
We define the rotation tensor $\omega_{\mu\nu}$ and the expansion tensor
$\theta_{\mu\nu}$ as follows,
\begin{align}
\label{2}
\omega_{\mu\nu} \equiv & \frac{1}{2} \left(U_{\mu;\alpha}h_\nu^\alpha
 -U_{\nu;\alpha}h_\mu^\alpha \right)\, , \\
\label{3}
\theta_{\mu\nu} \equiv &\frac{1}{2} \left(U_{\mu;\alpha}h_\nu^\alpha
+U_{\nu;\alpha}h_\mu^\alpha \right)\, ,
\end{align}
We also define the scalar expansion $\theta$ by
$\theta\equiv \theta_\mu^\mu={U^\mu}_{;\mu}$.
In the FRW space-time, whose metric is given by
\begin{equation}
\label{FRWmetric}
ds^2 = - dt^2 + a(t)^2 \sum_{i=1,2,3} \left( dx^i \right)^2 \, ,
\end{equation}
we find $\theta=3H$.
The shear tensor is given by
\begin{equation}
\label{4}
\sigma_{\mu\nu}\equiv \theta_{\mu\nu}
 - \frac{\theta}{3} h_{\mu\nu}=\theta_{\mu\nu}-Hh_{\mu\nu}\, .
\end{equation}
We should note that $\sigma_\mu^\mu=0$ by definition.
We may decompose the covariant derivative of $U_\mu$, $U_{\mu;\nu}$,
as follows,
\begin{equation}
\label{5}
U_{\mu;\nu}=\omega_{\mu\nu}+\sigma_{\mu\nu}+Hh_{\mu\nu}-A_\mu U_\nu\, ,
\end{equation}
where $A_\mu=\dot{U}_\mu=U^\alpha U_{\mu;\alpha}$ is the four-acceleration of the fluid.

By assuming the temperature $T$ is constant,
the energy-momentum tensor $T_{\mu\nu}$ of fluid is given by
\begin{equation}
\label{6}
T_{\mu\nu}=\rho U_\mu U_\nu + \left( p - \zeta \theta \right) h_{\mu\nu}
-2\eta \sigma_{\mu\nu}\, .
\end{equation}
Here
\begin{equation}
\label{7}
p_\mathrm{eff} \equiv p - \zeta \theta = p - 3\zeta H \, ,
\end{equation}
is the effective pressure, lower than $p$ because of the inequality $\zeta \geq 0$
which in turn is a consequence of thermodynamics.

\section{Gravitational Waves in a Viscous Fluid}

We now consider the energy-momentum tensor $\delta T_{\mu\nu}$ for
viscous fluid in (\ref{6}).
We should note that the energy density $\rho$ and the pressure $p$ depend on the
metric in general.
For simplicity, we assume,
\begin{equation}
\label{V1}
\delta \rho = \rho^{\mu\nu} \delta g_{\mu\nu}\, , \quad
\delta p = p^{\mu\nu} \delta g_{\mu\nu}\, .
\end{equation}
The shear viscosity $\eta$ and the bulk viscosity $\zeta$ may also depend
on the energy density $\rho$ and the pressure $p$.
In fact, it is often assumed to $\zeta \propto \rho^\lambda$, with a constant.
Because we have assumed that the energy density $\rho$ and the pressure $p$
should depend on metric as in (\ref{V1}), we may write
\begin{align}
\label{V2}
\delta\eta =& \eta^{(\rho)} \delta \rho + \eta^{(p)} \delta p
= \eta^{\mu\nu} \delta g_{\mu\nu}
\equiv \left( \eta^{(\rho)} \rho^{\mu\nu} + \eta^{(p)} p^{\mu\nu} \right)
\delta g_{\mu\nu}\, , \nn
\delta\zeta =& \zeta^{(\rho)} \delta \rho + \zeta^{(p)} \delta p
= \zeta^{\mu\nu} \delta g_{\mu\nu}
\equiv \left( \zeta^{(\rho)} \rho^{\mu\nu} + \zeta^{(p)} p^{\mu\nu} \right)
\delta g_{\mu\nu}\, .
\end{align}
In the FRW space-time (\ref{FRWmetric}) we may assume $\rho^{\mu\nu}$,
$p^{\mu\nu}$, $\eta^{\mu\nu}$, and $\zeta^{\mu\nu}$ only depend on the
cosmological time $t$ and do not depend on the spatial coordinates
$\left( x^i \right)$.
Because $U^\mu U_\mu = -1$, the variation of $U^\mu$ should satisfy the condition
\begin{equation}
\label{V3}
0 = 2 \left( \delta U^\mu \right) + U^\mu U^\nu \delta g_{\mu\nu}
= U^\mu \left( 2 g_{\mu\nu} \delta U^\nu + \delta g_{\mu\nu} U^\nu \right)\, ,
\end{equation}
which tells
\begin{equation}
\label{V4}
\delta U^\mu = - \frac{1}{2} g^{\mu\rho} \left( \delta g_{\rho\nu} U^\nu + l_\rho \right)\, .
\end{equation}
Here $l_\rho$ is a vector satisfying the condition $U^\mu l_\mu = 0$.
The vector $l_\rho$ will be determined later.
Eq.~(\ref{3}) also tells that
\begin{align}
\label{V5}
\delta \theta_{\mu\nu} =& \frac{1}{2} \left(\delta U_{\mu;\alpha}h_{\ \nu}^\alpha
 - \frac{1}{2}g^{\kappa\lambda}\left(
\nabla_\mu \delta g_{\alpha\lambda} + \nabla_\alpha \delta g_{\mu\lambda}
 - \nabla_\lambda \delta g_{\mu\alpha} \right) U_\kappa h_{\ \nu}^\alpha
 - U_{\mu;\alpha} U^\alpha \delta g_{\nu\xi} U^\xi
 - U_{\mu;\alpha} U^\alpha g_{\nu\xi} \delta U^\xi
 - U_{\mu;\alpha} \delta U^\alpha U_\nu \right. \nn
& \left. + \delta U_{\nu;\alpha}h_{\ \mu}^\alpha
 - \frac{1}{2}g^{\kappa\lambda}\left(
\nabla_\nu \delta g_{\alpha\lambda} + \nabla_\alpha \delta g_{\nu\lambda}
 - \nabla_\lambda \delta g_{\nu\alpha} \right) U_\kappa h_{\ \mu}^\alpha
 - U_{\nu;\alpha} U^\alpha \delta g_{\mu\xi} U^\xi
 - U_{\nu;\alpha} U^\alpha g_{\mu\xi} \delta U^\xi
 - U_{\nu;\alpha} \delta U^\alpha U_\mu \right) \, .
\end{align}
Therefore we find
\begin{align}
\label{V6}
\delta \theta =& - g^{\rho\mu} \delta g_{\mu\nu} g^{\nu\sigma} \theta_{\rho\sigma}
+ \delta U_{\mu;\alpha}h^{\mu\alpha}
 - \frac{1}{2}g^{\kappa\lambda}\left(
\nabla_\mu \delta g_{\alpha\lambda} + \nabla_\alpha \delta g_{\mu\lambda}
 - \nabla_\lambda \delta g_{\mu\alpha} \right) U_\kappa h^{\alpha\mu} \nn
& - g^{\mu\nu} U_{\mu;\alpha} U^\alpha \delta g_{\nu\xi} U^\xi
 - g^{\mu\nu} U_{\mu;\alpha} U^\alpha g_{\nu\xi} \delta U^\xi
 - U_{\mu;\alpha} \delta U^\alpha U^\mu \, .
\end{align}
and
\begin{align}
\label{V7}
\delta \sigma_{\mu\nu} = & \delta \theta_{\mu\nu}
 - \frac{1}{3} \delta \theta h_{\mu\nu}
 - \frac{\theta}{3} \left( \delta g_{\mu\nu} + \delta U_\mu U_\nu
+ U_\mu \delta U_\nu \right) \nn
=& \frac{1}{2} \left(\delta U_{\mu;\alpha}h_{\ \nu}^\alpha
 - \frac{1}{2}g^{\kappa\lambda}\left(
\nabla_\mu \delta g_{\alpha\lambda} + \nabla_\alpha \delta g_{\mu\lambda}
 - \nabla_\lambda \delta g_{\mu\alpha} \right) U_\kappa h_{\ \nu}^\alpha
 - U_{\mu;\alpha} U^\alpha \delta g_{\nu\xi} U^\xi
 - U_{\mu;\alpha} U^\alpha g_{\nu\xi} \delta U^\xi
 - U_{\mu;\alpha} \delta U^\alpha U_\nu \right. \nn
& \left. + \delta U_{\nu;\alpha}h_{\ \mu}^\alpha
 - \frac{1}{2}g^{\kappa\lambda}\left(
\nabla_\nu \delta g_{\alpha\lambda} + \nabla_\alpha \delta g_{\nu\lambda}
 - \nabla_\lambda \delta g_{\nu\alpha} \right) U_\kappa h_{\ \mu}^\alpha
 - U_{\nu;\alpha} U^\alpha \delta g_{\mu\xi} U^\xi
 - U_{\nu;\alpha} U^\alpha g_{\mu\xi} \delta U^\xi
 - U_{\nu;\alpha} \delta U^\alpha U_\mu \right) \nn
& - \frac{1}{3} \Bigl( - g^{\rho\eta} \delta g_{\eta\zeta} g^{\zeta\sigma}
\theta_{\rho\sigma} + \delta U_{\eta;\alpha}h^{\eta\alpha}
 - \frac{1}{2}g^{\kappa\lambda}\left(
\nabla_\eta \delta g_{\alpha\lambda} + \nabla_\alpha \delta g_{\eta\lambda}
 - \nabla_\lambda \delta g_{\eta\alpha} \right) U_\kappa h^{\alpha\eta} \nn
& - g^{\eta\zeta} U_{\eta;\alpha} U^\alpha \delta g_{\zeta\xi} U^\xi
 - g^{\eta\zeta} U_{\eta;\alpha} U^\alpha g_{\zeta\xi} \delta U^\xi
 - U_{\eta;\alpha} \delta U^\alpha U^\eta \Bigr)h_{\mu\nu}
 - \frac{\theta}{3} \left( \delta g_{\mu\nu} + \delta U_\mu U_\nu
+ U_\mu \delta U_\nu \right) \, .
\end{align}
Then $\delta T_{\mu\nu}$ has a rather complicated form,
\begin{align}
\label{V8}
\delta T_{\mu\nu}= & \delta \rho U_\mu U_\nu
+ \left( \rho + p - \zeta \theta \right) \delta U_\mu U_\nu
+ \left( \rho + p - \zeta \theta \right) U_\mu \delta U_\nu
+ \left( \delta p - \delta \zeta \theta - \zeta \delta \theta \right) h_{\mu\nu} \nn
& + \left( p - \zeta \theta \right) \delta g_{\mu\nu}
-2\delta \eta \sigma_{\mu\nu} -2\eta \delta \sigma_{\mu\nu} \, .
\end{align}
We will now need to simplify the formalism.

\subsection{Case of Bulk Viscosity}

First we ignore the shear viscosity by putting $\eta=0$.
We also investigate the propagation of the massless spin two mode, where
\begin{equation}
\label{V9}
\delta g_{it}=\delta g_{ti}=0\, , \quad \sum_{i=1,2,3} \delta g_{ii}=0 \, , \quad i=1,2,3\, .
\end{equation}
Then because $\delta g_{tt}=0$, we may assume $\delta U^\mu =0$.
Furthermore if we assume $\rho^{ij},\, p^{ij}\propto \delta^{ij}$, we find
$\delta \rho = \delta p = \delta \zeta = \delta \eta = 0$.
In the FRW space-time (\ref{FRWmetric}), we find
\begin{equation}
\label{V10}
\theta_{tt}=\theta_{ti}=\theta_{it}=U_{tt} = U_{t;i}=U_{i;t}=0\, , \quad
\theta_{ij} = U_{i;j}= a^2 H \delta_{ij}\, , \quad \theta = 3 H \, .
\end{equation}
Then by using (\ref{V5}), we find
\begin{equation}
\label{V11}
\delta \theta_{tt} = \delta \theta_{ti} = \delta \theta_{it} = 0\, , \quad
\delta \theta_{ij} = \frac{1}{2} \left(
2 \nabla_i \delta g_{jt} - \nabla_t \delta g_{ij} \right)
= - \frac{1}{2} \partial_t \delta g_{ij} \, , \quad
\delta \theta = 0\, .
\end{equation}
Then we obtain
\begin{equation}
\label{V12}
\delta T_{tt} = \delta T_{ti} = \delta T_{it} = 0 \, , \quad
\delta T_{ij} = \left( p - 3 H \zeta \right) \delta g_{ij} \, .
\end{equation}
By using (\ref{V9}) and (\ref{V12}), we see that Eqs.~(\ref{GRGW5}) and (\ref{GRGW7}) are
trivially satisfied.
On the other hand, Eq.~(\ref{GRGW6}) has the following form,
\begin{align}
\label{V13}
0 =& \frac{1}{2\kappa^2}\left( \frac{1}{2} \Box^{(0)} \delta g_{ij}
+ 2 \left( \dot H + H^2 \right) \delta g_{ij} \right)
+ \frac{1}{2} \left( p - 3 H \zeta \right) \delta g_{ij} \nn
=& \frac{1}{2\kappa^2}\left( \frac{1}{2}
\left( - \partial_t^2 \delta g_{ij} + a^{-2} \triangle \delta g_{ij} \right)
+ \left( 3 \dot H + 4 H^2 \right) \delta g_{ij} \right)
+ \frac{1}{2} \left( p - 3 H \zeta \right) \delta g_{ij} \, ,
\end{align}
Therefore the massless gravitational wave with spin two exists
even if the bulk viscosity is nonzero. This is consistent with Ref.~\cite{Goswami:2016tsu}.

\subsection{Case of Shear Viscosity}

We now consider the case when there exists only a shear viscosity, $\eta\neq 0$.
First we assume (\ref{V9}) and therefore $\delta U^\mu =0$.
Then (\ref{V7}) gives
\begin{align}
\label{V14}
\delta \sigma_{tt} =& 0 \, , \\
\delta \sigma_{ti} = & \delta \sigma_{it}
= \frac{1}{2} \left( - \frac{1}{2}g^{\kappa\lambda}\left(
\nabla_t \delta g_{\alpha\lambda} + \nabla_\alpha \delta g_{t\lambda}
 - \nabla_\lambda \delta g_{t\alpha} \right) U_\kappa h_{\ i}^\alpha \right) =0 \, , \\
\delta \sigma_{ij} = & \frac{1}{2} \left( - \frac{1}{2}g^{\kappa\lambda}\left(
\nabla_i \delta g_{\alpha\lambda} + \nabla_\alpha \delta g_{i\lambda}
 - \nabla_\lambda \delta g_{i\alpha} \right) U_\kappa h_{\ j}^\alpha
 - \frac{1}{2}g^{\kappa\lambda}\left(
\nabla_j \delta g_{\alpha\lambda} + \nabla_\alpha \delta g_{j\lambda}
 - \nabla_\lambda \delta g_{j\alpha} \right) U_\kappa h_{\ i}^\alpha \right) \nn
& - \frac{1}{3} \left( - g^{\rho\eta} \delta g_{\eta\zeta} g^{\zeta\sigma}
\theta_{\rho\sigma} - \frac{1}{2}g^{\kappa\lambda}\left(
\nabla_\eta \delta g_{\alpha\lambda} + \nabla_\alpha \delta g_{\eta\lambda}
 - \nabla_\lambda \delta g_{\eta\alpha} \right) U_\kappa h^{\alpha\eta}
\right)h_{ij}
 - \frac{\theta}{3} \delta g_{ij} \nn
= & - \frac{1}{2} \partial_t \delta g_{ij} - H \delta g_{ij} \, .
\end{align}
Then Eqs.~(\ref{GRGW5}) and (\ref{GRGW7}) are trivially satisfied, again.
On the other hand, Eq.~(\ref{GRGW6}) has the following form,
\begin{align}
\label{V15}
0 =& \frac{1}{2\kappa^2}\left( \frac{1}{2} \Box^{(0)} \delta g_{ij}
+ 2 \left( \dot H + H^2 \right) \delta g_{ij} \right)
+ \frac{1}{2} \left( p - 3 H \zeta \right) \delta g_{ij}
 - 2 \eta \left( - \frac{1}{2} \partial_t \delta g_{ij} - H \delta g_{ij} \right) \nn
=& \frac{1}{2\kappa^2}\left( \frac{1}{2}
\left( - \partial_t^2 \delta g_{ij} + a^{-2} \triangle \delta g_{ij} \right)
+ \left( 3 \dot H + 4 H^2 \right) \delta g_{ij} \right)
+ \frac{1}{2} \left( p - 3 H \zeta \right) \delta g_{ij}
 - 2 \eta \left( - \frac{1}{2} \partial_t \delta g_{ij} - H \delta g_{ij} \right) \, ,
\end{align}
This equation tells that also in this case a massless spin two gravitational wave exists.
We should note, however, the presence of the term $\eta \partial_t \delta g_{ij}$.
If $\eta>0$, Eq.~(\ref{V15}) tells that the gravitational wave is enhanced.
On the other hand, if $\eta<0$, the term $\eta \partial_t \delta g_{ij}$ expresses
a dissipation of the gravitational wave. The absorbed wave energy is transformed into heat.

\section{Cosmological speculations}

Because bulk viscosity generates effectively a negative pressure (positive tensile stress),
a fluid endowed with bulk viscosity will lead to an accelerated expansion.
Although bulk viscosity and shear viscosity have different physical backgrounds
(as mentioned the shear viscosity is usually far the greatest among them), it is natural
to include them both in the formalism.

For the recently observed gravitational waves
\cite{Abbott:2016blz,Abbott:2016nmj,Abbott:2017vtc,Abbott:2017oio,
Abbott:2017gyy,TheLIGOScientific:2017qsa},
the distances between the sources and the earth are about a few hundreds Mpc.
Because no dissipation or enhancement of the gravitational waves has been
observed we find the following constraint on the shear viscosity $\eta_0$
in the present universe,
\begin{equation}
\label{Cos1}
\left| \kappa^2 \eta_0 \right| \ll \left( 10^3\, \mathrm{Mpc}\right)^{-1}\, ,
\end{equation}
even if the accelerated expansion of the universe is generated by the
viscous fluid. Let us express this constraint in physical units:
as 1 Mpc=$3.086\times 10^{22}~$m
and 1 s=$3\times 10^8~$m in geometric units, the right hand side of Eq.~(\ref{Cos1})
becomes $10^{-17}~$s$^{-1}$ in physical units.
Then, since $\kappa^2=1.87\times 10^{-26}~$m/kg, we can write the constraint as
\begin{equation}
\eta_0 \ll 5\times 10^8~ \rm{ Pa~s}. \label{constraint1}
\end{equation}
In case that the viscous fluid has electromagnetic interaction, the fluid may absorb,
reflect, or radiate the light, which may affect the luminosity distance and
the above constraint could be changed.
We now assume, however, that the viscous fluid does not have electromagnetic interaction
as in the many models of the dark energy or dark matter.
Then the luminosity distance is not changed in our model and under the assumption,
the above constraint (\ref{Cos1}) could be valid.
Another point we note is that if the dark energy is represented by the viscous fluid,
the present bulk viscosity $\zeta$ satisfies
\begin{equation}
\label{Cos2}
\frac{1}{\kappa^2 \zeta_0} \sim \frac{1}{H} \sim 10^4\, \mathrm{Mpc}\, .
\end{equation}
or in physical units
\begin{equation}
\zeta_0 \sim 5\times 10^7~\rm{Pa~s}\, .
\label{constraint2}
\end{equation}
It is notable that this value of $\zeta_0$ is only one or two orders of magnitude greater than
the value derived from comparison with experiments. We return to this point in Sec. VII.

On the other hand, if the inflation in the early universe could be generated by the
viscous fluid, we may expect that the shear viscosity could be large.
Then the primordial gravitational wave may have been absorbed into the viscous fluid.
Then even if the primordial gravitational wave may not be observed in the future
experiments, this might be due to the viscous fluid.
In fact, the period of the inflation could be estimated to be $10^{-34}$ sec.
$\sim 10^{19}\,\mathrm{eV}=10^{10}\, \mathrm{GeV}$.
Here we have used the following natural unit,
\begin{equation}
\label{unit}
1\,\mathrm{s} = 1.5192674\times 10^{15}\,\mathrm{eV}^{-1}\, ,\quad
1\,\mathrm{m} = 5.0677307\times 10^{6}\,\mathrm{eV}^{-1}\, , \quad
1\,\mathrm{kg} = 5.6095886\times 10^{35}\,\mathrm{eV}\, .
\end{equation}
This tells that if the primordial gravitational wave is detected, we find
\begin{equation}
\label{Cos3}
\left| \eta \kappa^2 \right| \ll 10^{19}\,\mathrm{eV}=10^{10}\, \mathrm{GeV} \, .
\end{equation}
On the other hand, if the scale of the inflation is the GUT scale
$\sim 10^{15}\, ,\mathrm{GeV}$, we find
\begin{equation}
\label{Cos4}
\left| \kappa^2 \zeta \right| \sim H \sim 10^{- 19 + 2\times15}
= 10^{11}\, \mathrm{GeV}\, .
\end{equation}
By comparing (\ref{Cos3}) and (\ref{Cos4}), if
$\left| \zeta \right| \sim \left| \eta \right|$, Eq.~(\ref{Cos3}) could not be
satisfied and therefore the primordial gravitational wave may be absorbed into
the viscous fluid.
Then if the primordial gravitational wave will be detected by the future observation,
we obtain the constraint (\ref{Cos3}) but if the inflation was generated by the viscous fluid,
the primordial gravitational wave might not be detected.

\section{First Example: Evolution of Gravitational Waves near the Big Rip Singularity}

If the universe is filled with a viscous fluid, a Big Rip singularity may appear.
In this section, we investigate the behavior of the gravitational field close to the singularity.

\subsection{Cosmology in the Presence of a Viscous Fluid}

Before investigating the propagation and the evolution of the gravitational wave near the Big
Rip singularity, we will review the evolution of the viscous fluid by
neglecting other components like dark matter and ordinary matter.
The energy-momentum tensor in (\ref{6})
gives the following FRW equations
\begin{equation}
\label{Cos5}
\frac{3}{\kappa^2} H^2 = \rho \, , \quad
 - \frac{1}{\kappa^2} \left( 3H^2 + 2\dot H \right) = p - 3\zeta H \, ,
\end{equation}
which gives the conservation law,
\begin{equation}
\label{Cos6}
0 = \dot \rho + 3 H \left( \rho + p - 3\zeta H \right) \, .
\end{equation}
We note that the shear viscosity does not contribute to the
background evolution.
We also assume $p \propto \rho$ and $\zeta \propto \rho^\lambda$ with a
constant $\lambda$ as follows,
\begin{equation}
\label{Cos7}
p = w \rho \, , \quad \zeta= \tilde\zeta_0 \rho^\lambda \, .
\end{equation}
Here $w$ is the equation of state (EoS) parameter and $\tilde\zeta_0$ is a constant.
(The subscript zero in this section refers to an arbitrary starting point near the singularity,
not to the present time as above.)
Then by using the first FRW equation in (\ref{Cos5}), we can rewrite (\ref{Cos6})
as follows,
\begin{equation}
\label{Cos8}
0 = \dot \rho + \kappa \left(w +1 \right) 3^{\frac{1}{2}} \rho^{\frac{3}{2}}
 - 3 \kappa^2 \tilde\zeta_0 \rho^{\lambda+1} \, ,
\end{equation}
Especially in case of $\lambda=\frac{1}{2}$, which is often chosen, by redefining the
EoS parameter as
\begin{equation}
\label{Cos9}
w_\mathrm{eff} = w - 3^{\frac{1}{2}} \kappa \tilde\zeta_0\, ,
\end{equation}
Eq.~(\ref{Cos8}) can be rewritten as,
\begin{equation}
\label{Cos10}
0 = \dot \rho + \kappa \left(w_\mathrm{eff} +1 \right) 3^{\frac{1}{2}} \rho^{\frac{3}{2}}
= \dot \rho + 3H \left(w_\mathrm{eff} +1 \right) \rho\, ,
\end{equation}
which is nothing but the conservation law for a standard perfect fluid with EoS
parameter $w_\mathrm{eff}$.
Therefore even if $w>-1$, in case $w_\mathrm{eff} <-1$, the fluid becomes effectively
phantom and generates a Big Rip singularity, where $H$ and $a$ behave as
\begin{equation}
\label{Cos11}
H = \frac{h_0}{t_s - t} \, , \quad a = a_0 \left( t_s - t \right)^{-h_0}\, , \quad
h_0 \equiv - \frac{2}{3 \left( w_\mathrm{eff} + 1 \right)} > 0 \, .
\end{equation}
Hence there occurs a Big Rip singularity at $t=t_s$.

\subsection{Propagation Near the Big Rip Singularity}

By using (\ref{V15}), we consider the propagation and evolution of gravitational
waves near the singularity in (\ref{Cos11}). The equation of state (\ref{Cos7}) is adopted.
Because we are considering spatially flat background, we consider a plane wave,
where $\delta g_{ij} \propto \e^{i\bm{k}\cdot\bm{x}}$.
We also assume that $\zeta$, like $\eta$, is proportional to
$\rho^{\frac{1}{2}} \propto H$ and we write
\begin{equation}
\label{Cos12}
\eta = \frac{\tilde\eta_0}{\kappa^2} H \, .
\end{equation}
Then Eq.~(\ref{V15}) has the following form,
\begin{align}
\label{Cos13}
0 \sim & \frac{1}{2\kappa^2}\left( \frac{1}{2}
\left( - \partial_t^2 \delta g_{ij} - a_0^{-2} \left( t_s - t \right)^{2h_0}k^2 \delta g_{ij} \right)
+ \frac{3h_0 + 4 h_0^2}{\left( t_s - t \right)^2}\delta g_{ij} \right)
+ \frac{3w_\mathrm{eff} h_0^2}{2 \kappa^2 \left( t_s - t \right)^2} \delta g_{ij} \nn
& - \frac{2 \tilde\eta_0 h_0}{\kappa^2 \left(t_s - t\right)} \left( - \frac{1}{2} \partial_t \delta g_{ij}
 - \frac{h_0}{t_s - t} \delta g_{ij} \right) \nn
\sim & \frac{1}{4\kappa^2}\left\{ - \partial_t^2 \delta g_{ij}
+ \frac{2 \tilde\eta_0 h_0}{t_s - t} \partial_t \delta g_{ij}
+ \frac{3h_0 + 4 h_0^2 + 6 w_\mathrm{eff} h_0^2 + 8 \tilde\eta_0 h_0^2}
{\left( t_s - t \right)^2}\delta g_{ij} \right\} \, ,
\end{align}
Because the last equation in (\ref{Cos13}) is homogeneous,
the solution can be obtained by assuming
\begin{equation}
\label{Cos14}
\delta g_{ij} \propto \left( t_s - t \right)^\alpha \, .
\end{equation}
Here $\alpha$ is a constant.
Then Eq.~(\ref{Cos13}) can be rewritten as an algebraic equation,
\begin{align}
\label{Cos15}
0 =& \alpha^2 + \left( 2 \tilde\eta_0 h_0-1\right) \alpha
 - \left( 3h_0 + 4 h_0^2 + 6 w_\mathrm{eff} h_0^2 + 8 \tilde\eta_0 h_0^2\right) \nn
=& \alpha^2 + \left( 2 \tilde\eta_0 h_0-1\right) \alpha
 - \left( - h_0 - 2 h_0^2 + 8 \tilde\eta_0 h_0^2\right)
\, .
\end{align}
Here we have deleted $w_\mathrm{eff}$ by using the definition of $h_0$ in (\ref{Cos11}).
The solution of (\ref{Cos15}) is given by
\begin{align}
\label{Cos16}
\alpha = \alpha_\pm \equiv& \frac{1}{2} \left( - 2 \tilde\eta_0 h_0 + 1 \pm
\sqrt{ \left( 2 \tilde\eta_0 h_0-1\right)^2
+ 4 \left( - h_0 - 2 h_0^2 + 8 \tilde\eta_0 h_0^2\right)} \right) \nn
=& \frac{1}{2} \left( - 2 \tilde\eta_0 h_0 + 1 \pm
\sqrt{ 4 h_0^2 \tilde\eta_0^2 + \left( - 4 h_0 + 32 h_0^2 \right) \tilde\eta_0
+ 1 - 4 h_0 - 8 h_0^2} \right) \, .
\end{align}
When $\tilde\eta_0=0$, that is, there is no shear viscosity, $\alpha_\pm$ is always
positive because we are assuming $h_0>0$ and therefore $\delta g_{ij}$ goes to
vanish near the singularity.
However in case $- h_0 - 2 h_0^2 + 8 \tilde\eta_0 h_0^2>0$, that is,
\begin{equation}
\label{Cos17}
\tilde\eta_0 > \frac{1}{4} + \frac{1}{8 h_0}\, ,
\end{equation}
$\alpha_-$ becomes negative and therefore $\delta g_{ij}$ will be enhanced
near the singularity.
We note that $\tilde\eta_0$ is positive if Eq.~(\ref{Cos17}) is satisfied.

We now consider more details.
The determinant $D$ of the quadratic algebraic equation (\ref{Cos15}) with respect to
$\alpha$ is given by
\begin{equation}
\label{Cos18}
D = 4 h_0^2 \tilde\eta_0^2 + \left( - 4 h_0 + 32 h_0^2 \right) \tilde\eta_0
+ 1 - 4 h_0 - 8 h_0^2 \, .
\end{equation}
We note that $D>0$ when $h_0<\frac{1}{6}$ or even if $h_0>\frac{1}{6}$ when
\begin{equation}
\label{Cos19}
\tilde\eta_0 > \frac{1}{2 h_0} - 4 + \sqrt{ \frac{1}{4 h_0^2} - \frac{4}{h_0} + 16
 - \frac{1}{4h_0^2} + \frac{1}{h_0} + 2 }
= \frac{1}{2 h_0} - 4 + \sqrt{ - \frac{3}{h_0} + 18 } \quad
\mbox{or} \quad
\tilde\eta_0 < \frac{1}{2 h_0} - 4 - \sqrt{ - \frac{3}{h_0} + 18 } \, .
\end{equation}
On the other hand, $D<0$ if $h_0>\frac{1}{6}$ and
\begin{equation}
\label{Cos20}
\frac{1}{2 h_0} - 4 - \sqrt{ - \frac{3}{h_0} + 18 }
< \tilde\eta_0 < \frac{1}{2 h_0} - 4 + \sqrt{ - \frac{3}{h_0} + 18 } \, .
\end{equation}
If $D>0$, $\alpha_\pm$ are real but if $D<0$, $\alpha_\pm$ becomes complex and
$\left( \alpha_+ \right)^\dagger = \alpha_-$, therefore the amplitude of the gravitational
wave oscillates but if $\tilde\eta_0 > \frac{1}{2h_0}$, the amplitude decreases and
if $\tilde\eta_0 < \frac{1}{2h_0}$, the amplitude increases.
We note that the solution of the algebraic equation
\begin{equation}
\label{Cos21}
\frac{1}{4} + \frac{1}{8 h_0} = \frac{1}{2h_0} \, ,
\end{equation}
is $h_0=\frac{3}{2}$.
Then when $h_0>\frac{3}{2}$, $\frac{1}{4} + \frac{1}{8 h_0} > \frac{1}{2h_0}$ and
when $h_0<\frac{3}{2}$, $\frac{1}{4} + \frac{1}{8 h_0} > \frac{1}{2h_0}$.
On the other hand, we find
\begin{equation}
\label{Cos22}
\frac{1}{2 h_0} - 4 - \sqrt{ - \frac{3}{h_0} + 18 } < \frac{1}{2h_0} \, ,
\end{equation}
for positive $h_0$ and the algebraic equation
\begin{equation}
\label{Cos23}
\frac{1}{2 h_0} - 4 + \sqrt{ - \frac{3}{h_0} + 18 } = \frac{1}{2h_0} \, ,
\end{equation}
has the solution $h_0=\frac{3}{2}$, again.
Therefore when $h_0>\frac{3}{2}$,
$\frac{1}{2 h_0} - 4 + \sqrt{ - \frac{3}{h_0} + 18 } > \frac{1}{2h_0}$ and
when $\frac{3}{2}>h_0>\frac{1}{6}$,
$\frac{1}{2 h_0} - 4 + \sqrt{ - \frac{3}{h_0} + 18 } < \frac{1}{2h_0}$.
We also find that as long as $h_0>\frac{1}{6}$,
$\frac{1}{2 h_0} - 4 - \sqrt{ - \frac{3}{h_0} + 18 } < \frac{1}{4} + \frac{1}{8 h_0}$.
The solution of the algebraic equation
\begin{equation}
\label{Cos24}
\frac{1}{2 h_0} - 4 + \sqrt{ - \frac{3}{h_0} + 18 } = \frac{1}{4} + \frac{1}{8 h_0} \, ,
\end{equation}
is only $h_0=\frac{3}{2}$.
Then when $h_0\neq \frac{3}{2}$,
$\frac{1}{2 h_0} - 4 + \sqrt{ - \frac{3}{h_0} + 18 } < \frac{1}{4} + \frac{1}{8 h_0}$.

Then the above results can be summarized as follows,
\begin{itemize}
\item Case $0<h_0<\frac{1}{6}$:
\begin{itemize}
\item When $\tilde\eta_0 < \frac{1}{4} + \frac{1}{8 h_0}$, the amplitude of the gravitational wave
is monotonically decreasing.
\item When $\tilde\eta_0 > \frac{1}{4} + \frac{1}{8 h_0}$, in addition to the mode where
the amplitude of the gravitational wave is monotonically decreasing, there appears another
mode where the amplitude of the gravitational wave is monotonically increasing.
\end{itemize}
\item Case $\frac{1}{6} < h_0 < \frac{3}{2}$:
\begin{itemize}
\item When $\tilde\eta_0 < \frac{1}{2 h_0} - 4 - \sqrt{ - \frac{3}{h_0} + 18}$ or
$\frac{1}{2 h_0} - 4 + \sqrt{ - \frac{3}{h_0} + 18 } < \tilde\eta_0 < \frac{1}{4} + \frac{1}{8 h_0}$,
the amplitude of the gravitational wave is monotonically decreasing.
\item When $\frac{1}{2 h_0} - 4 - \sqrt{ - \frac{3}{h_0} + 18 }
< \tilde\eta_0 < \frac{1}{2 h_0} - 4 + \sqrt{ - \frac{3}{h_0} + 18 }$, the amplitude of
the gravitational wave is oscillating and decreasing.
\item When $\tilde\eta_0 > \frac{1}{4} + \frac{1}{8 h_0}$, in addition to the mode where
the amplitude of the gravitational wave is monotonically decreasing, there appears another
mode where the amplitude of the gravitational wave is monotonically increasing.
\end{itemize}
\item Case $h_0>\frac{3}{2}$:
\begin{itemize}
\item When $\tilde\eta_0 < \frac{1}{2 h_0} - 4 - \sqrt{ - \frac{3}{h_0} + 18}$ or
$\frac{1}{2 h_0} - 4 + \sqrt{ - \frac{3}{h_0} + 18 } < \tilde\eta_0 < \frac{1}{4} + \frac{1}{8 h_0}$,
the amplitude of the gravitational wave is monotonically decreasing.
\item When $\frac{1}{2 h_0} - 4 - \sqrt{ - \frac{3}{h_0} + 18} < \tilde\eta_0 < \frac{1}{2h_0}$,
the amplitude is decreasing with oscillation.
\item When $\frac{1}{2h_0} < \tilde\eta_0 < \frac{1}{2 h_0} - 4 + \sqrt{ - \frac{3}{h_0} + 18}$,
the amplitude is increasing with oscillation.
\item When $\tilde\eta_0 > \frac{1}{4} + \frac{1}{8 h_0}$, in addition to the mode where
the amplitude of the gravitational wave is monotonically decreasing, there appears another
mode where the amplitude of the gravitational wave is monotonically increasing.
\end{itemize}
\end{itemize}

We may compare the above result with the case that there is no shear viscosity, $\eta=0$.
Then Eq.~(\ref{Cos16}) reduces to
\begin{equation}
\label{Cos25}
\alpha = \frac{1}{2} \left( 1 \pm
\sqrt{ 1 - 4 h_0 - 8 h_0^2} \right) \, .
\end{equation}
Therefore if $1 - 4 h_0 - 8 h_0^2>0$, that is,
\begin{equation}
\label{Cos26}
0< h_0 < \frac{-2 + \sqrt{5}}{8} \, ,
\end{equation}
the amplitude of the gravitational wave decreases monotonically.
If $1 - 4 h_0 - 8 h_0^2<0$, that is,
\begin{equation}
\label{Cos27}
h_0 > \frac{-2 + \sqrt{5}}{8} \, ,
\end{equation}
the amplitude of the gravitational wave decreases with oscillation.
Then we find that there is no case in which the amplitude is enhanced, as in the case of
$\eta \neq 0$.

The Big Rip singularity is very violent, it might look that the linear analyses might be
broken but the non-linear effects become important when the amplitude of the
gravitational wave becomes large but as we have shown, in many cases,
the amplitude decreases and therefore in these cases, we may neglect
the non-linear effect.
In other cases, the amplitude increases and the non-linear effect and also quantum
effects may become important.

\section{Second Example: Gravitational Waves in the Lepton Era}

As second example we will go to an opposite extreme and apply the above formalism to
the early universe, specifically to the end point of the lepton era.
The reason for this choice is the following.

The lepton era is characterized by a temperature drop from $10^{12}\,$K (100\,MeV)
to $10^{10}\,$K (1\,MeV).
In this period the particles present were essentially photons, neutrinos, and electrons,
together with their antiparticles.
Transfer of momentum took place among the particles because the relativistic
and non-relativistic species decreased in temperature following different powers
of the scale factor: relativistic particles decreased as $T\propto a^{-1}$ while the
non-relativistic ones decreased as $T \propto a^{-2}$.
The bulk viscosity arose
because the photons, electrons and the $(e, \mu, \tau)$ leptons had short mean
free paths compared with those of the neutrinos. The maximum momentum transfer,
taking place at the instant of neutrino decoupling, $10^{10}\,$K, marked the maximum
of the bulk viscosity.

The theory of viscosities in this very early region was worked out in the extensive
study \cite{hoogeveen86}, with use of the relativistic Boltzmann equation.
Related works are \cite{leeuwen86} and \cite{Caderni:1977rd}.
A recent detailed treatment, emphasizing the connection with particle physics,
is given in \cite{Husdal:2016ofa}.
The general behavior of the bulk viscosity in the lepton era is that it is varying very much,
by 7-8 orders of magnitude, when regarded as a function of $T$.
The reason for this is the influential, but rapidly decaying, $\tau$ and $\mu$ mesons.
The maximum value at neutrino decoupling is \cite{Husdal:2016ofa}
\begin{equation}
\label{1a}
\zeta \approx 10^{22}\, \mathrm{Pa\, s}\, .
\end{equation}
An important point is that this value can be connected with the present value $\zeta_0$
of the bulk viscosity, via the relation
\begin{equation}
\label{2a}
\zeta= \zeta_0\left(\frac{\rho}{\rho_0}\right)^\lambda\, ,
 \end{equation}
with $\lambda$ a constant (here again we let subscript zero refer to the present time).

Much work has been done on the phenomenological level, by comparing measured values
of $H=H(z)$ with predictions from the FRW equations, in order to determine the value
of $\zeta_0$ (for a recent review, see \cite{Brevik:2017msy}). From our own
investigations~\cite{Normann:2016jns,Normann:2016zby}, it turned out that
one could restrict $\zeta_0$ to the interval
\begin{equation}
\label{3a}
10^4\,\mathrm{Pa\,s} \leq \zeta_0 \leq 10^6\,\mathrm{Pa\,s}\, .
\end{equation}
It should be noted that this equation is roughly comparable with Eq.~(\ref{constraint2}) above.

Now, combining Eq.~(\ref{1a}) with the scaling (\ref{2a}), putting $\lambda=1/2$,
we ended up with the value $\zeta_0= 10^5\,$Pa\,s, thus in the logarithmic middle of
the interval (\ref{3a}). This coincidence, which could hardly have been foreseen,
gives support to the bulk viscosity numbers above. We here mention also the formula
for the energy density as a function of temperature~\cite{Husdal:2016ofa},
\begin{equation}
\label{4a}
\rho=\frac{\pi^2}{30}g_{*\rho}T^4\, ,
\end{equation}
with $g_{*\rho}$ being the effective degree of freedom for $\rho$.
At the instant of neutrino decoupling, $g_{*\rho} \approx 11$.

Now move on to the shear viscosity $\eta$.
From Fig.~8 in \cite{Husdal:2016ofa} we read off
\begin{equation}
\label{5a}
\eta \approx 10^{25}\,\mathrm{Pa\,s}
\end{equation}
at neutrino decoupling.
The shear viscosity is thus much greater than the bulk viscosity,
as is usual in hydrodynamics.
It should be borne in mind that this is a theoretical result, derived from
the relativistic Boltzmann equation; there is not an experimental link to the present
value $\eta_0$ as was the case for the bulk viscosity.
The reason is of course the avoidance of spatial isotropy as soon as shear viscosity
is concerned. It might seem natural nevertheless to assume simply that the scaling
relation (\ref{2a}) holds approximately also for the shear viscosity,
\begin{equation}
\label{6a}
\eta=\eta_0\left( \frac{\rho}{\rho_0}\right)^\lambda\, .
\end{equation}
($\tilde\eta_0 = \kappa \sqrt{\frac{3}{\rho_0}}$ in (\ref{Cos12}) when
$\lambda=\frac{1}{2}$.)
Then, we can calculate the present shear viscosity $\eta_0$, using (\ref{6a}) and (\ref{2a}), as
\begin{equation}
\label{7a}
\eta_0=\frac{\eta \zeta_0}{\zeta} \approx 10^8~\rm{Pa~s}
\end{equation}
for any constant $\lambda$, assuming $ \zeta_0 \approx 10^{5}\,\mathrm {Pa\,s}$.

Armed with this formalism, we can now express $\delta T_{\mu\nu}$ in (\ref{V8})
in such a way that the variations of the viscosity coefficients are carried back to
the variations of the energy density,
\begin{align}
\label{8a}
\delta \zeta =& \frac{\lambda \zeta_0}{\rho_0^\lambda}\rho^{\lambda-1}\delta \rho \, , \\
\label{9a}
\delta \eta =& \frac{\lambda \eta_0}{\rho_0^\lambda}\rho^{\lambda-1}\delta \rho\, .
\end{align}
Moreover, for the pressure we can assume, as usual in the radiation epoch,
\begin{equation}
\label{10a}
p=\frac{1}{3}\rho\, .
\end{equation}

We now consider the gravitational waves.
This case is simplified, since
$\delta \rho=\delta p=\delta \zeta=\delta \eta=0$.
Then by using (\ref{V15}), one gets
\begin{equation}
\label{11a}
0 = \frac{1}{2\kappa^2}\left( \frac{1}{2}(-\partial_t^2\delta g_{ij}
+a^{-2}\Delta \delta g_{ij})+(3\dot{H}+4H^2)\delta g_{ij} \right)
+\frac{1}{6}(\rho-9H\zeta)\delta g_{ij}+2\eta\left( \frac{1}{2}\partial_t \delta g_{ij}
+H\delta g_{ij}\right)\, .
\end{equation}
As $\zeta \ll \eta$ at the time of neutrino decoupling, we can omit the $\zeta$ term
and the equation reduces to
\begin{equation}
\label{12a}
\frac{1}{2\kappa^2}\left( \frac{1}{2}(-\partial_t^2\delta g_{ij}+a^{-2}\Delta \delta g_{ij})
+(3\dot{H}+4H^2)\delta g_{ij} \right)
+\frac{1}{6}(\rho+12\eta H)\delta g_{ij}+\eta \partial_t \delta g_{ij}=0\, .
\end{equation}
We may note that at this instant, in physical units,
\begin{equation}
\label{13a}
\rho c^2 \approx 10^{25}\,\mathrm{J/m^3}\, ,
\end{equation}
and moreover $t=1\,$s so that $H=1/2t=1/2\,$s$^{-1}$.
Thus the influence from the viscous term $12\eta H$ is, perhaps surprisingly,
of the same order of magnitude as the energy density term $\rho c^2$.

The equation (\ref{12a}) is complicated, since it contains both $\rho$ and $H$.
We can eliminate $\rho$ by making use of the FRW equation
\begin{equation}
3H^2=\kappa^2 \rho\, ,
\label{14a}
\end{equation}
and by taking $\lambda=1/2$ we can express the shear viscosity as
\begin{equation}
\label{15a}
\eta=\frac{\eta_0}{\kappa}\sqrt{\frac{3}{\rho_0}}\, H\, .
\end{equation}
Equation (\ref{12a}) then takes the form
\begin{equation}
\label{16a}
\frac{1}{2}(-\partial_t^2\delta g_{ij}+a^{-2}\Delta \delta g_{ij})+3\dot{H}\delta g_{ij}
+\left( 5+4\kappa \eta_0\sqrt{\frac{3}{\rho_0}}\right)H^2\delta g_{ij}
+2\kappa {\eta_0}\sqrt{\frac{3}{\rho_0}}H\partial_t\delta g_{ij}=0\, .
\end{equation}
Introducing, as usual,
\begin{equation}
\label{17a}
\delta g_{ij}=a^2 h_{ij}\, , \quad |h_{ij}| \ll 1\, ,
\end{equation}
we then have
\begin{equation}
\label{18a}
 -\partial_t^2(a^2 h_{ij})+\Delta h_{ij}+6\dot{H}a^2h_{ij}
+2\left( 5+4\kappa \eta_0\sqrt{\frac{3}{\rho_0}}\right)H^2a^2h_{ij}
+ 4\kappa {\eta_0}\sqrt{\frac{3}{\rho_0}}\, H\partial_t(a^2h_{ij})=0\, .
\end{equation}
It is advantageous to get information about the numerical magnitudes from viscosities here.
Going over to physical units we obtain, inserting $G=6.67\times 10^{-11}\,$Nm$^2$/kg$^2$,
$\eta_0 =10^8\,$ Pa\,s, $\rho_0 \approx 10^{ -26}\,$kg/m$^3$,
\begin{equation}
\label{19a}
4\kappa \eta_0\sqrt{\frac{3}{\rho_0}} \rightarrow
\frac{4\sqrt{4\pi G}}{c}\eta_0\sqrt{\frac{3}{\rho_0c^2}} \approx 3.0\, .
\end{equation}
This number is remarkably close to one.
Thus Eq.~(\ref{18a}) can be written as
\begin{equation}
\label{20a}
 -\partial_t^2(a^2 h_{ij})+\Delta h_{ij}+6\dot{H}a^2h_{ij}
+16H^2a^2h_{ij} + 3 H\partial_t(a^2h_{ij})=0\, .
\end{equation}
This is the governing equation for gravitational waves on a time-dependent background.
Recall what is the basis for this equation: it relies upon the connecting formulas (\ref{2a})
and (\ref{6a}) for the viscosities together with the assumption $\lambda=1/2$, and also
on the numerical values for $\zeta_0$ and $\eta_0$. Otherwise, there is no direct link to
the neutrino decoupling time in (\ref{20a}) except through the values for $a$ and $H$.
The equation consequently has a quite a general value; it is applicable to any instant
within the radiation dominated epoch.

We may process this equation further by introducing conformal time $\tau$ via $dt=ad\tau$,
letting a prime henceforth mean derivative with respect to $\tau$.
We also replace $h_{ij}$ with the quantity $\mu_{ij}=ah_{ij}$.
Then, with ${\mathcal{H}}=a^\prime/a=aH$ we have
\begin{equation}
\label{21a}
\partial_t^2(a^2h_{ij})=\frac{1}{a}\left( {\mu^\prime}_{ij}+{\mathcal{H}}{\mu^\prime}_{ij}
 -{\mathcal{H}}^2\mu_{ij}+\frac{a^{\prime\prime}}{a}\mu_{ij}\right)\, ,
\end{equation}
and by transforming the other terms in (\ref{20a}) similarly we can write the equation as
\begin{equation}
\label{22a}
{\mu^{\prime\prime}}_{ij}-2{\mathcal{H}}{\mu^\prime}_{ij}
 -\left( 8{\mathcal{H}}^2+5\frac{a^{\prime\prime}}{a}\right)\mu_{ij}-\Delta \mu_{ij}=0\, .
\end{equation}
Going over to Fourier space, taking $\mu_{ij}$ to vary as $\e^{i(kr-\omega \tau)}$,
we obtain the dispersion relation
\begin{equation}
\label{23a}
\omega^2=k^2-2i{\mathcal{H}}k-\left( 8{\mathcal{H}}^2
+5\frac{a^{\prime\prime}}{a}\right)=0\, .
\end{equation}
We assume that the gravitational wave moves on a slowly varying background
(the varying metric).
Then, $k^2$ can be taken to be much larger than ${\mathcal{H}}^2$ or
$a^{\prime\prime}/a$, and the relation reduces to
\begin{equation}
\label{24a}
\omega=k\sqrt{1-2i{\mathcal{H}}/k} \approx k(1-i{\mathcal{H}}/k)\, .
\end{equation}
This wave is dispersive; it corresponds formally to an electromagnetic wave
in a weakly absorbing medium whose complex refractive index is
$n^\prime =1+i{\mathcal{H}}/\omega$.
The decay constant with respect to position is thus ${\mathcal{H}}^{-1}$,
so that the amplitude for a wave with constant $\omega$ decays as
$\e^{-{\mathcal{H}}r}$.

The following point ought to be noted. It might seem as if memory about viscosity
is completely lost in the dispersion relation (\ref{24a}), as only $\mathcal{H}$ appears.
However, $\mathcal{H}$ comes from the scale factor, calculated from the FRW equations
in which the viscosity makes a contribution.
Also, recall that $h_{ij}$ (or $\mu_{ij}$) as solved from the governing equation (\ref{20a})
for waves on a slowly varying background relies upon equation (\ref{19a}) which is a
viscosity-related approximation.

It is of interest to compare the above results with other studies. 
That means first of all a comparison with the developments in Ref.~\cite{Goswami:2016tsu}, 
as our approaches are basically related. 
It is at once striking to observe that our estimate (\ref{7a}) for $\eta_0$ agrees so well with 
the critical shear viscosity obtained by these authors: they found 
$\eta_\mathrm{crit}=\rho_\mathrm{crit}H_0^{-1}=4.38\times 10^8~$Pa~s, with 
$\rho_\mathrm{crit}=3H_0^2 M_{Pl}^2$, and the present shear viscosity was written as 
$\eta_0 = Q\eta_\mathrm{crit}$ with $Q$ a factor of order unity. 
This agreement is encouraging, recalling that our estimate for $\eta_0$ in (\ref{7a}) was based 
on an analogy with the scaling law for the {\it bulk} viscosity and thus quite different from the argument 
leading to the value given in (\ref{7a}).

A second point is to compare our amplitude decay law $\e^{-{\cal H}r}$ with that derived in 
\cite{Goswami:2016tsu}. 
Observing that the physical distance $L$ corresponding to the distance $r$ is $L=ar$, 
these authors derived the decay law for the amplitude to be essentially $\e^{-3QLH_0}$, with 
$Q= O(0.1)H_0^{-1}$. This was found to correspond to about 25\% amplitude decay during the travel 
to the detector (central value of $L$ being 410 Mpc). 
Again, we find a satisfactory agreement, as our decay law reads $\e^{-HL/c}$ in dimensional units. 
There are of course numerically large differences in the two cases, as we are focusing on the lepton region 
for which $H=1/(2t)$. 
For instance, at neutrino decoupling $t=1~$s, and so the wave amplitude gets attenuated by a factor of 
$e$ by propagation over a distance of only $6\times 10^5~$km.

Gravitational wave damping is a many-faceted phenomenon, which can be caused by various processes. 
Instead of describing it as viscosity-driven as we have done, one may alternatively consider it via 
the development of the anisotropic stress tensor when influenced by the free steaming neutrinos. 
This is the approach of Weinberg {\it et al.} \cite{Flauger:2017ged,Weinberg:2003ur}. 
The basic governing equation is then the collisionless Boltzmann equation. 
The damping due to these processes was found to be quite significant, as the B-B polarization multipole 
coefficients were found to be reduced by about 35\% because of the free-streaming neutrinos. 
A generalization of this method was carried out by Baym {\it et al.} \cite{Baym:2017xvh}, including collisions 
and particles of finite mass. 
Some years earlier, Durrer and Kahniashvili \cite{Durrer:1997ta} studied the CMB anisotropies caused 
by gravitational waves.

One may finally ask: is there a direct link between the present calculation and the physics of 
neutrino decoupling? 
We think there is such a link, although an indirect one. The basic ingredient behind the calculation 
of the shear and bulk viscosities is after all the relativistic Boltzmann equation, applied to the lepton epoch 
which is in turn taken to be composed of the three lepton families $(e, \mu, \tau)$, 
in addition to the neutrinos. The most important viscosity coefficient is the bulk viscosity $\zeta$, 
in view of the assumed spatial isotropy. As shown explicitly in Fig.~9 in Ref.~\cite{Goswami:2016tsu}, 
for example, the value of $\zeta$ varies very much. 
For each of the lepton species, $\zeta$ at first increases steeply, as $T^{-5}$, and thereafter dies out rapidly. 
Thus the contribution from the $\tau$ dies out for $T>10^{12}~$K, that from the $\mu$ dies out for 
$T> 10^{11}~$K, while that from the $e$ is both much bigger and lasts longer, until about $10^9~$K. 
The maximum value of $\zeta$ is about $10^{24}~$ Pa~s. Our theory implies that we link neutrino decoupling 
to the approximate maximum of the bulk viscosity, which is physically in accordance with the circumstance 
that the momentum transfer between neutrinos and the electromagnetically interacting particles is then 
at maximum. When the viscosity disappears there in no particle system to interact with for the neutrinos; 
that is, they decouple.

\section{Concluding Remarks}

It should be borne in mind that basic approach to describing the cosmic fluid is after all the one of
statistical mechanics, where the Boltzmann equation and the Chapman-Enskog expansion enables
one to calculate the viscosity coefficients \cite{chapman52,vincenti65}.
A condition for this method to work is that the free mean path is much shorter than the wavelengths.
It is not evident that this condition is satisfied for the cosmic fluid.
The dominant kind of matter is dark matter, and little is known about its microscopic structure,
even at the present time.
And if we go to the extreme epochs, either the very early, or the very late, universe, even less is
known.
In spite of these concerns we have found it reasonable to make use of a viscous model for the
cosmic fluid, mainly because of its simplicity and its applicability in almost all areas of fluid
mechanics.
There might be a microstructure in the dark matter making the viscous approach quite permissible.
Notably, the viscous approach has been followed by a large group of investigators, among them the
recent Ref.~\cite{Goswami:2016tsu}.

We derived the governing equation for gravitational waves propagating in a fluid with bulk viscosity
$\zeta$ and shear viscosity $\eta$ to be as given in (\ref{V15}). In Sec. VI we took the equation of
state to be the conventional $p=w\rho$, and took the bulk and shear viscosities to be given
respectively by Eqs.~(\ref{Cos7}) and (\ref{Cos12}).
With the forms for $H$ and $a$ given in (\ref{Cos11}) we could then discuss whether the gravitational
wave either increases, or decreases, near the Big Rip singularity.
This was our first example.
Our second example in Sec. VII was taken from the early universe, specifically the lepton era when
the temperature dropped from $10^{12}~$K to $10^{10}~$K.
The motivation for this choice was that at the end of this era the microscopic bulk viscosity is
at maximum. In this case we took only the bulk viscosity into account, in view of the commonly
accepted spatial isotropy of the cosmic fluid.
We found the governing equation for gravitational waves propagating on the underlying medium,
assumed to be ``slowly varying'', and found the wave to behave essentially as an absorbing medium in
optics where the decay constant was the inverse of the conformal Hubble parameter
${\mathcal{H}}=aH$.

Our results for the damping were found to be in quite good agreement with those obtained recently 
in Ref.\cite{Goswami:2016tsu}
 - a paper based upon the cosmic viscosity model - similarly as ours.

\section*{Acknowledgments}

We are indebted to S.~D.~Odintsov for the discussions and collaboration at early stage.
This work is supported (in part) by
MEXT KAKENHI Grant-in-Aid for Scientific Research on Innovative Areas ``Cosmic
Acceleration'' No. 15H05890 (S.N.) and the JSPS Grant-in-Aid for
Scientific Research (C) No. 18K03615 (S.N.).

\end{document}